\pdfoutput=1

\documentclass[11pt]{article}

\usepackage[]{acl}

\usepackage{times}
\usepackage{latexsym}
\usepackage{graphicx}
\usepackage[export]{adjustbox}
\usepackage{bm}
\usepackage{amssymb}
\usepackage{float}
\usepackage{booktabs}
\usepackage{multirow}
\usepackage{subfigure}
\usepackage{makecell}
\usepackage{amsmath} 
\usepackage{url}
\usepackage{times}
\usepackage{latexsym}
\usepackage{colortbl}
\usepackage{graphicx}
\usepackage{caption}
\usepackage{subcaption}

\usepackage[T1]{fontenc}

\usepackage[utf8]{inputenc}

\usepackage{microtype}

\usepackage{inconsolata}
\usepackage[T1]{fontenc}

\usepackage[utf8]{inputenc}

\usepackage{microtype}

\usepackage{inconsolata}

\title{FinVerse: An Autonomous Agent System for Versatile Financial Analysis}

\author{Siyu An$^1$, Qin Li$^1$, Junru Lu$^2$, Di Yin$^1$ and Xing Sun$^1$ \\
$^1$Tecent YouTu Lab, $^2$University of Warwick \\
$^1$\texttt{\{siyuan,jemmali,endymecyyin,winfredsun\}@tencent.com} \\
$^2$\texttt{junru.lu@warwick.ac.uk}}

\begin{document}
\maketitle
\begin{abstract}
With the significant advancements in cognitive intelligence driven by LLMs, autonomous agent systems have attracted extensive attention. 
Despite this growing interest, the development of stable and efficient agent systems poses substantial practical challenges. In this paper, we introduce FinVerse, a meticulously crafted agent system designed for a broad range of financial topics. FinVerse integrates over 600 financial APIs, enabling access to more accurate and extensive financial information compared to generalist agents.
To enhance financial information processing capabilities, FinVerse is equipped with an embedded code interpreter, enabling the execution of complex data analysis tasks with precision and efficiency. 
Our work includes an empirical comparison of several LLMs in driving FinVerse. Specifically, we propose our own scheme for training LLMs using SFT to optimize LLM performance within FinVerse.  
Recognizing the scarcity of specialized datasets to build LLMs for agents, we have  constructed a dataset and plan to make it open-source, providing a valuable resource for peer application developers. 
The demo video has been released on YouTube\footnote{\url{https://www.youtube.com/watch?v=sk8L9_Wv7J4}}.

\end{abstract}


\section{Introduction}
With the explosive growth of cognitive intelligence brought by recent advancements in large language models (LLMs) \citep{brown2020language, ouyang2022training, touvron2023llama, yang2023baichuan}, the boundaries of artificial intelligence have been dramatically expanded.
Although LLMs are powerful in a wide range of tasks, they still face limitations in certain areas, such as acquiring factual knowledge, solving math problems, and accessing real-time information. 
To address these inadequacies, previous studies\citep{schick2023toolformer, nakano2022webgpt, qin2023webcpm} have highlighted that the integration of auxiliary tools, such as web search engines and calculators, could significantly augment the capabilities of LLMs.

\begin{figure}[t]
    \centering
    \includegraphics[width=0.9\linewidth]{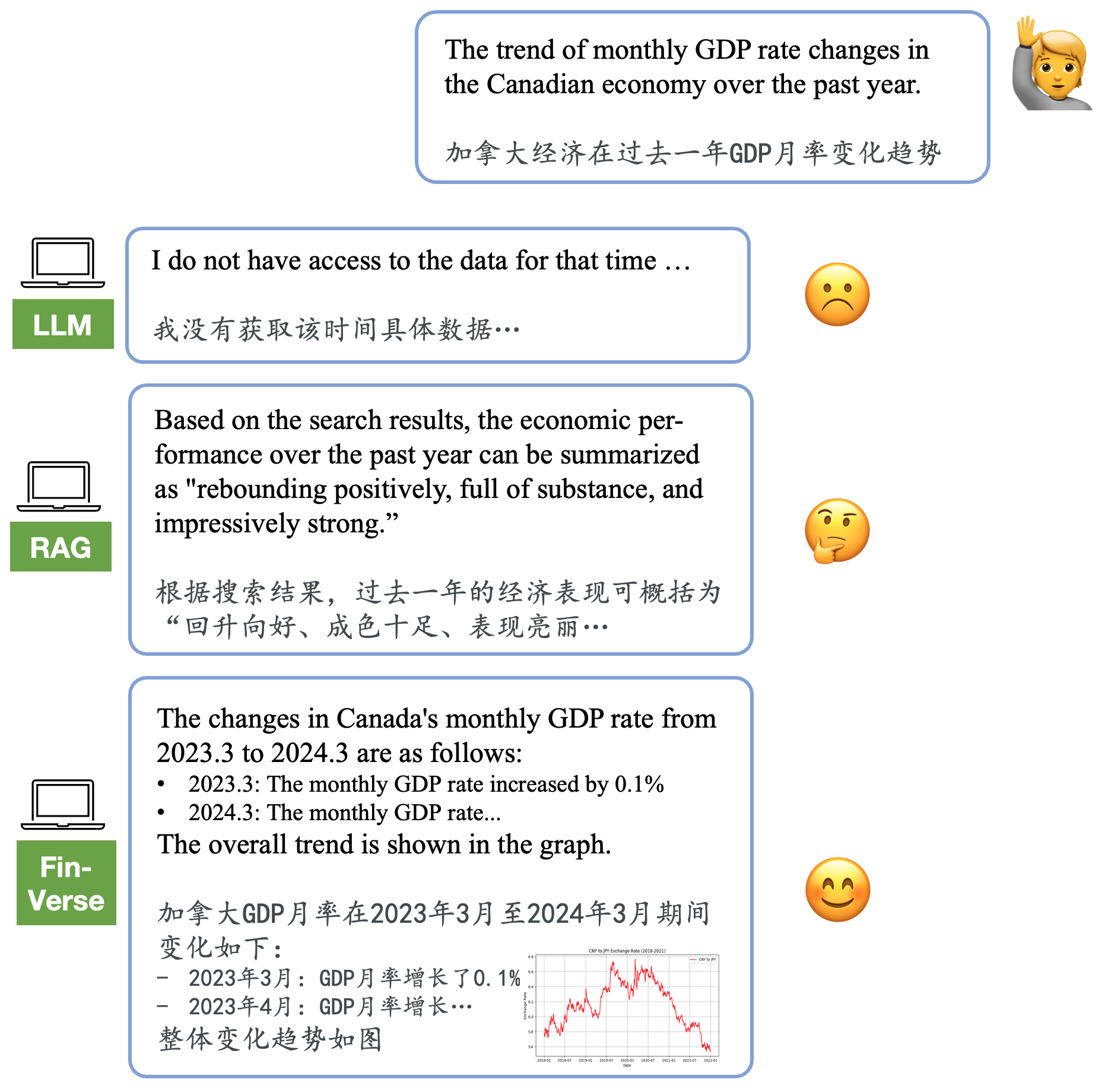}
    \centering
    \caption{A comparative illustration of the anticipated outputs across various systems}
    \label{fig:sim_pic}
\end{figure}

The integration of LLMs with external plugins is typically exemplified by Retrieval-Augmented Generation (RAG). RAG primarily compensates for the deficiency of LLMs in acquiring external information by introducing document retrieval. RAG is widely applicable and has demonstrated stable performance across many tasks\citep{baumann2024combining, chen2023benchmarking, gao2024retrievalaugmented}. However, the limitations of retrieval in attempting to acquire more granular data also reveal the constraints of RAG when it comes to addressing certain complex problems. To address such issues, our designed FinVerse allows LLMs to access over 600 specialized APIs in the financial sector to acquire richer data, and integrates the code interpreter for analysis, thereby yielding more comprehensive results.In Fig.\ref{fig:sim_pic}, we showcase a comparative illustration of the anticipated outputs across various systems, thereby underscoring the enhanced complexity and depth of the results delivered by FinVerse.

Another form for LLMs to utilize plugins is autonomous agent. Technologies such as ReAct \citep{yao2022react}, Reflexion \citep{shinn2023reflexion}, and Memory \citep{hatalis2023memory} have been shown to improve LLMs' efficiency in task planning and tool utilization. Consequently, there has been a growing interest in the development of autonomous agents systems\citep{wang2023survey, wu2023autogen, talebirad2023multi} that synergistically integrate these enhancements, with LLMs serving as the core component. 
Current efforts are primarily focused on developing the LLMs capabilities as the center of agents in general domains \citep{qin2023toolllm, zeng2023agenttuning, pan2024kwaiagents}. However, recently literatures indicate the performance of LLMs in this pivotal role is marked by notable deficiencies \citep{xie2024travelplanner}. To mitigate the shortcomings of the LLMs from a systemic perspective while maximizing its capabilities, the system is designed to enhance overall stability and performance as effectively as possible.

\begin{figure*}[h]
    \centering
    \includegraphics[width=1\linewidth]{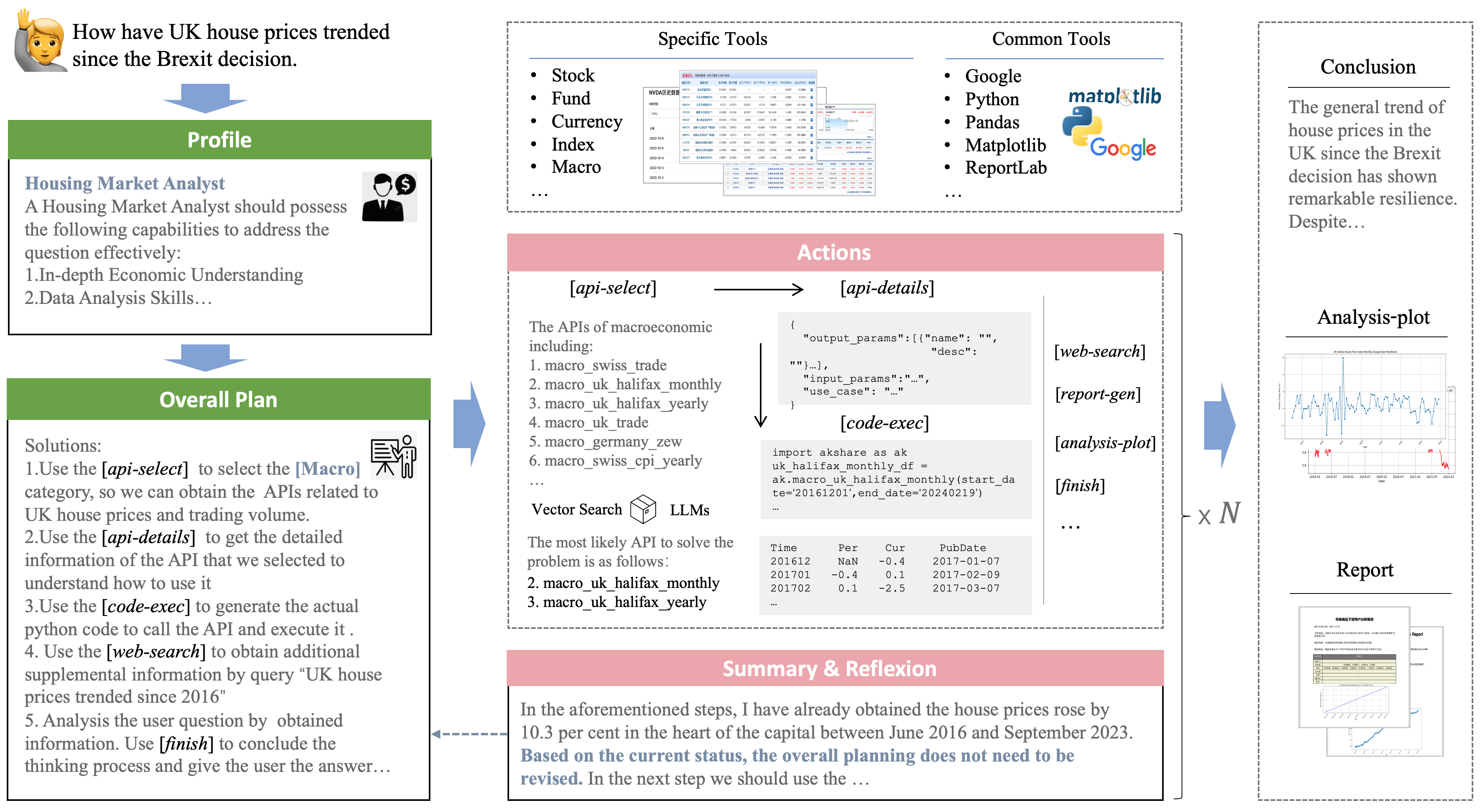}
    \centering
    \caption{The overall architecture of FinVerse}
    \label{fig:main}
\end{figure*}

In this paper, we introduce a finical agent system called FinVerse which achieves excellent performance on general financial topics including stocks, funds, macroeconomics, etc. 
In FinVerse, we reference ReAct process for task scheduling, utilizing a pipeline as shown in Fig \ref{fig:main}. For the user's initial question, we first establish the most appropriate Profile and Overall Plan to tackle the task, ensuring that the system continuously retains memory of this information throughout all subsequent steps.
Subsequently, guided by the Profile and Overall Plan, the agent system progressively generates various actions by LLMs. After each action, the Summary\&Reflexion module allows LLMs to evaluate the invocation outcomes. If LLMs identify any errors or missteps in the task flow, it adapts and tweaks the Overall Plan, granting the system the capability for pathway correction.

To ensure the model could accurately locate and employ financial APIs, we have first identified 642 APIs across 10 categories to access real-time financial data. Unlike typical tool invocation methods where LLMs select the APIs and parameters in the form of slot filling, in our approach LLMs interpret the API documents to generate the code embedded with the API callings, thus enhancing the extensibility. For the accurate invocation of financial APIs for data retrieval, we have established three actions: \textit{api-select}, \textit{api-details}, and \textit{code-exec}. In \textit{api-select}, we take LLMs to predict the API category and employ vector search to control the number of APIs recalled. In \textit{api-details}, we source the API document based on its name to serve as part of the prompt for the LLMs. Finally, during the \textit{code-exec} phase, the LLMs draft code based on the API document and the sub-task in current step, with the execution results fed back to the LLMs for subsequent operations.

Due to the insufficient capabilities of open-source LLMs, currently most agents rely on commercial LLMs with robust capabilities, such as GPT4, to ensure smooth operations. To address this, we have developed a specialized dataset for Supervised Fine Tuning(SFT) that emphasizes four crucial sub-tasks: Overall Plan, Action Taking, Summary \& Reflexion and Code Writing. Experiments show the SFT process has greatly enhanced the capabilities of open-source LLMs to act as a substitute in the agent system.

In summary, our contributions are as follows:

(1) We present an agent framework focused on financial topics to produce comprehensive outcomes. The experience is transferable to other contexts, offering broader applicability.

(2) We fine-tuning open-source LLMs in four sub-tasks and introduce the practices to empower LLMs with the capability as the agent core.

(3) To compensate for the scarcity of specialized datasets for agents, we have compiled a dataset that will be made open source upon paper acceptance.


\section{Approaches}



\subsection{System Architecture}


The agent system is composed of LLMs and the surroundings for tools using. Our agent system architecture, intended for performing a range of financial tasks, incorporates a suite of domain-specific and general-purpose tools. 
For domain-specific tasks, we have integrated an extensive collection of financial data APIs, totaling 642, which are categorised into ten distinct domains of financial intelligence. 
To support a broad spectrum of functions, our infrastructure also includes a Docker containerized environment. This space facilitates the execution of analysis code via SDKs, and the use of a general search engine to acquire additional information. For high-performance inference, the LLMs are deployed on a Linux platform empowered by 8 NVIDIA Tesla A100 GPUs.

The primary workflow of FinVerse is illustrated in Figure \ref{fig:main}. When receiving a query, the system activates LLMs to initially identify a suitable profile including character descriptions and abilities to address the issue. Concurrently, LLMs outline the sequence of steps for resolution, which we named Overall Plan. During this phase, both the Profile and the Overall Plan serve as agent meta and are retained in memory for subsequent operations. 

In the Overall Plan, the steps take the action we designed previously. An action refers to a unit of behavior encapsulated by our custom-designed agent. For example, in Figure\,\ref{fig:main}, \textit{web-search} represents the action of invoking a general search engine. Specifically, we designed \textit{api-select}, \textit{api-details} and \textit{code-exec} to work together to facilitate the invocation of specialized tools. \textit{finish} is also set as an action to signal the system when to conclude the task.
After each action, we have implemented a module named Summary \& Reflexion to process the information retrieved from tool invocations. If LLMs find the results from a tool's invocation are unsatisfactory, it is required to amend the Overall Plan accordingly. 
On the other hand, if the results are as anticipated, LLMs summarize the findings of the preceding tool invocation and commit these insights to memory, thereby adhering to the roadmap established by the Overall Plan for subsequent actions.


\subsection{Key Capabilities}
In this section, we analysis the major abilities of LLMs within FinVerse and introduce the prompts.

\paragraph{Agent Meta}
When the agent system receives a query, it generates a Profile and an Overall Plan to address the issue, which is a standard operation within the agent framework. In FinVerse, the Overall Plan takes the specific actions we defined step by step, and the Overall Plan could be modified in subsequent steps. In Fig. \ref{fig:plan-roles} we present the prompt template to generate agent meta.

\paragraph{Tool Search}

The specific tools are the major routes for data acquisition in FinVerse. The difficulty is typically requires to find the appropriate one from a search space with hundreds of APIs available. To address this challenge, generally there are two strategies.
The first one is hierarchical search, which is akin to a human-like method of looking up specific one in numerous APIs. Firstly we leverage LLMs to infer the relevant category based on the task context. Then within the APIs in chosen category, LLMs determines the suitable one based on its API description. This method utilizes inherent API categorization and offers strong interpretability. However the drawback lies in the potential slowdown of LLMs reasoning due to lengthy inputs when a category contains a large number of APIs.
The second strategy involves converting API description texts into embedding vectors and employing vector matching to identify the appropriate one. While this method disregards inherent features such as API categories and may yield lower accuracy when only the top few results are selected.
To leverage the strengths of both methods, we combine them in our usage. Initially, the LLMs select the most suitable API category based on the user's query and the agent meta. 
For categories with an extensive number of specific APIs, such as stocks, we adopt embedding search to retrieve API candidates, and take LLMs to select the suitable one.

In practical use, we encapsulate two actions over the original specific APIs: \textit{api-select} and \textit{api-details} as descriped in Fig.\ref{fig:main}. The \textit{api-select} takes the category of the specific APIs and a description of the problem, returning a set of candidate specific APIs. When the number of candidates is large, embedding search is used to limit the number of candidates to $M$. When the LLMs confirm $m$\text{-}th API, \textit{api-details} retrieves information such as the API's input parameters, output parameters, and usage examples, which is useful for LLMs to write codes for employing this specialized API.



\paragraph{Code Writing}
After acquiring the API documentation, we directly utilize LLMs to generate code that calls the API based on the current task plan and API documentation. Instead of having LLMs select tools and fill in slots, Code Writing enhances the transferability of our solution and facilitates handling scenarios with numerous APIs. This code is then executed in a containerized environment and the results of the code execution are returned to LLMs for further understanding and analysis. 
Another advantage is that after we obtain data using the SDK, if further analysis is required based on the returned data, such as calculating maximum and minimum values or creating plots with matplotlib, it can also be executed via code in this step. This significantly enhances the capability of FinVerse to handle complex tasks, as demonstrated in our demo video.
Therefore, FinVerse can provide users with richly illustrated and detailed analytical results.
The prompt template is presented in Fig. \ref{fig:code-exec}.

\paragraph{Summary\&Reflexion}
This module fulfills two main roles. Firstly, the action might return an extensive amount of detailed data. To facilitate the preservation and utilization of the data, we leverage LLMs to extract insights that are valuable for resolving the task. This approach conserves space in the prompt templates in later processes, typically considered as short-term memory. Secondly, the LLMs continuously monitor the workflow to ensure it is proceeding as expected. In FinVerse, this module is executed through a single call to LLMs.



\section{Experiments}
\subsection{Experimental Setup}

\paragraph{Dataset} 
For specific tools, we captured real-time financial data by leveraging financial APIs from an open-source financial data interface library known as AkShare\footnote{\url{https://github.com/akfamily/akshare}}. Our data process included parsing the API documentation, from which professional annotators selected and rigorously checked the API's availability. 
The statistical details of candidate APIs after cleaning are presented in Table \ref{tab:tools description}. 
To facilitate our project, we've compiled a dataset featuring a diverse array of financial questions from real-world scenarios. 
To ensure there was no data leakage, We performed deduplication on the questions to ensure no data leakage and got 14107 questions. Finally, the questions are divided into training, evaluation and testing sets in rate 8:1:1.


For the purpose of training custom-built LLMs, it is common practice to utilize state-of-the-art LLMs, such as GPT-4, to orchestrate the entire workflow. This approach involves capturing and preserving the inputs and outputs generated by the LLMs and distill their capabilities into a form that can be utilized for further training. To more precisely refine the model's abilities, we have developed four specific tasks, each designed to enhance particular aspects of the model's performance.


\subsection{Experimental Results}

\subsubsection{Task-specific Performance}

To assess FinVerse's ability to accurately identify the appropriate API among hundreds of options, we designed experiments involving LLMs and vector retrieval techniques. We utilize BGE \citep{chen2024bge} for vector retrieval. For the LLMs, we take Baichuan2-13b-chat \cite{yang2023baichuan} and Qwen-14b-chat \cite{bai2023qwen} as our base models. The efficacy of these models was quantified by measuring both the "all right" and "all wrong" percentage rates within the Top-5 and Top-10 retrieval results. These findings have been systematically catalogued in Table \ref{tab: tool search}, with FV-BC2-13b denoting results derived from the fine-tuned Baichuan2-13b-chat model, and FV-QW-14b representing the model based on Qwen-14b-chat. The results indicate that combining vector retrieval with an LLM-based selection substantially enhances performance, particularly in the Top-10 recall, which demonstrates the efficacy of our method.
\begin{table}
\centering
\scalebox{0.7}
{
    \begin{tabular}{c|c|c|c|c}
    \toprule
    \multirow{2}*{\textbf{Method}} & \multicolumn{2}{c}{\textit{Top5}} & \multicolumn{2}{|c}{\textit{Top10}}\\
    \cline{2-5}
    {} &  \textbf{all right} & \textbf{all wrong} & \textbf{all right} & \textbf{all wrong}\\
    \hline
    BGE & 31.73 & 21.16 & 44.30 & 14.34 \\
    FV-BC2-13b & 34.96 & 28.10 & 35.86 & 27.70 \\
    FV-QW-14b & 29.52 & 34.99 & 29.86 & 34.37 \\
    BGE + FV-BC2-13b & \textbf{69.55} & \textbf{2.47} & \textbf{83.51} & \textbf{2.41} \\
    BGE + FV-QW-14b & 51.85 & 10.41 & 63.31 & 10.28 \\
    \bottomrule
    \end{tabular}
}
\caption{\label{tab: tool search}
The efficacy of various tool selection methods.
}
\end{table}

\begin{table}
\centering
\scalebox{0.8}
{
    \begin{tabular}{c|cccc}
    \toprule
    \textbf{LLM} &  \textbf{Plan} & \textbf{Action} & \textbf{Code} & \textbf{Summary}\\
    \hline
    GPT4 & \underline{98.51} & \underline{99.01} & \underline{90.55} & \underline{97.50} \\
    Baichuan2-13b & 3.36 & 2.97 & 13.39 & 23.33 \\
    Qwen-14b & 6.42 & 4.95 & 11.02 & 29.17 \\
    KW-BC2-13b & 47.76 & 64.36 & 8.67 & 21.67 \\
    KW-QW-14b & 51.49 & 56.44 & 14.96 & 17.50 \\
    FV-BC2-13b & \textbf{94.78} & \textbf{97.03} & 71.65 & 68.33 \\
    FV-QW-14b & 92.54 & 94.06 & \textbf{76.38} & \textbf{75.83} \\
    \bottomrule
    \end{tabular}
}
\caption{\label{tab: loss ablation study}
The performance of different LLMs on the main tasks of FinVerse with SFT alignment.
}
\end{table}
In Table~\ref{tab: loss ablation study}, we present the performance of various LLMs on the four central tasks within FinVerse: Overall Planning, Action Taking, Code Writing, and Summary\&Reflexion. For the Code Writing task, effectiveness is measured by the passing rate of the generated code and the accuracy of its outputs. For the remaining tasks, outcome correctness is determined through extensive human evaluations. The models subjected to comparison include GPT-4, Baichuan2-13b-chat, Qwen-14b-chat, and two adaptations from KwaiAgents \citep{pan2024kwaiagents}, namely ``kagentlms-baichuan2-13b-mat'' and ``kagentlms-qwen-14b-mat'', both of which are fine-tuned versions of their respective base models. Our models enhanced via SFT are denoted with the ``FV'' prefix. The table reveals a notable performance discrepancy of agent systems, particularly with certain open-source models that have tens of billions of parameters, when benchmarked against GPT-4. For fair comparison, we employed a standardized testing prompt template. Given that the fine-tuning of our ``FV'' models was tailored to this prompt template during training, this specialized alignment may account for the observed significant performance enhancements.


\subsubsection{End-to-End Performance}

Evaluating the overall performance of an agent system is a challenging task. For FinVerse, we have set three perspectives to assess it.
 
\subparagraph{Robustness}
To measure the agent robustness we verify its ability to accurately conclude a task within an appropriate number of steps $L$. 
In fact, we have observed crashes of agent systems are common in practice. The interrupts are mainly caused by three reasons: 1) The task could still not be solved in steps $L$. 2) The output format could not be parsed correctly. 3) The prompt length exceeds the LLMs configuration unexpectedly.
Consequently, we have established robustness as the foundational benchmark for our agent framework.

In Table \ref{tab:Main Results}, we compare the robustness of different agent systems. As a generalist agent, AutoGPT\footnote{\url{github.com/Significant-Gravitas/AutoGPT}} is not tailored to any specific task. We set it as one of our baselines, configured by default with web-search and GPT4 capabilities. In terms of robustness, we find that AutoGPT is comparatively more stable, yet it still fails to resolve some queries within a constrained number of steps. Additionally, we assess FinVerse's performance across various LLMs and discover that FV-QW-14b delivered the highest robustness. Besides, 
the number of specific tools do not influence the robustness of FinVerse.

\begin{table}[t]
\centering
\scalebox{0.7}{
    \begin{tabular}{l|*5l}
    \toprule
    \textbf{Agents} & \textbf{Best} & \textbf{Helpful} & \textbf{Robust} & \textbf{Freq-Tool} & \textbf{Freq-LLM}\\ 
    \hline
    AutoGPT & 32.13  & 1.32 & \underline{94.32} & 10.21 & 10.78 \\
    \hline
    FV-GPT4 & \underline{76.04}  & \underline{2.51} & 89.32 & 5.03 & 9.84 \\
    FV-ToolLlama & 2.18  & 0.16 & 11.33 & 1.25 & 2.47 \\ 
    FV-KW-QW & 4.28  & 0.24 & 7.52 & 1.45 & 1.20\\
    FV-KW-BC & 4.75  & 0.35 & 5.41 & 1.12 & 1.04\\
    \hline
    FV-BC2-13b$^{\dag}$ & 39.18 & 1.81 & 87.89 & 5.03 & 10.06 \\
    FV-BC2-13b & \textbf{55.15} & 2.06 & 82.22 & 5.38 & 9.73 \\
    FV-QW-14b$^{\dag}$ & 29.90 & 1.36 & 67.01 & 5.14 & 10.28\\
    FV-QW-14b & 53.61 & \textbf{2.09} & \textbf{88.92} & 6.02 & 10.75 \\
    \bottomrule
\end{tabular}}
\caption{Experimental results of FinVerse with max steps \textit{L=10}. $^{\dag}$ indicates the agents with only web-tools.}
\label{tab:Main Results}
\end{table}

\subparagraph{Helpfulness}
To assess the agent's final outcome in terms of its helpfulness for the task, the answers are evaluated with 0-3 scores provided by 5 financial experts. 
Specifically, a score of 3 represents an ideal answer, characterized by thorough and accurate data references and culminations, signifying that the agent is operating in exactly the way we set. 
A score of 2 denotes a response that is correct though of lower quality, typically including various web sources and links, yet the substantial model did not aptly discern the valuable data therein. 
A 1-point score signifies incorrect conclusions, typically indicating a misjudgment in the agent's operational process. In this scenario, the user can still salvage some useful information from the proceedings
A score of 0 indicates that the agent was unable to successfully conclude its procedures. The helpfulness score reflects the weighted outcome of the aforementioned scoring assessment.

From the Table \ref{tab:Main Results}, we observe that agents utilizing specialized tools significantly outperform those relying solely on web tools in both ``Best'' and ``Helpfulness'' scores. Additionally, aside from GPT-4, our models enhanced through SFT, achieve better performance within the FinVerse context.

\subparagraph{Efficiency}
Efficiency is another necessary criterion for assessing the usability of our agent system. In this paper we evaluate the efficiency of an agent by the calling frequency of LLMs and tools. In Table \ref{tab:Main Results}, we found that when employing open-source models directly, the frequency of agent calls for tools and LLMs is relatively low. This is attributed to the models' inability to adequately adapt to FinVerse, resulting in premature cessation of tasks. Moreover, we observed that most open-source LLMs can only accurately understand and utilize web-search tools; they tend to fail when it comes to invoking specialized tools. 
In addition, we discover that our SFT models, FV-BC2-13b and FV-QW-14b, reach a better balance between efficacy and efficiency.

\section{Demo Showcase}

\begin{figure}[h]
    \centering
    \includegraphics[width=1.02\linewidth]{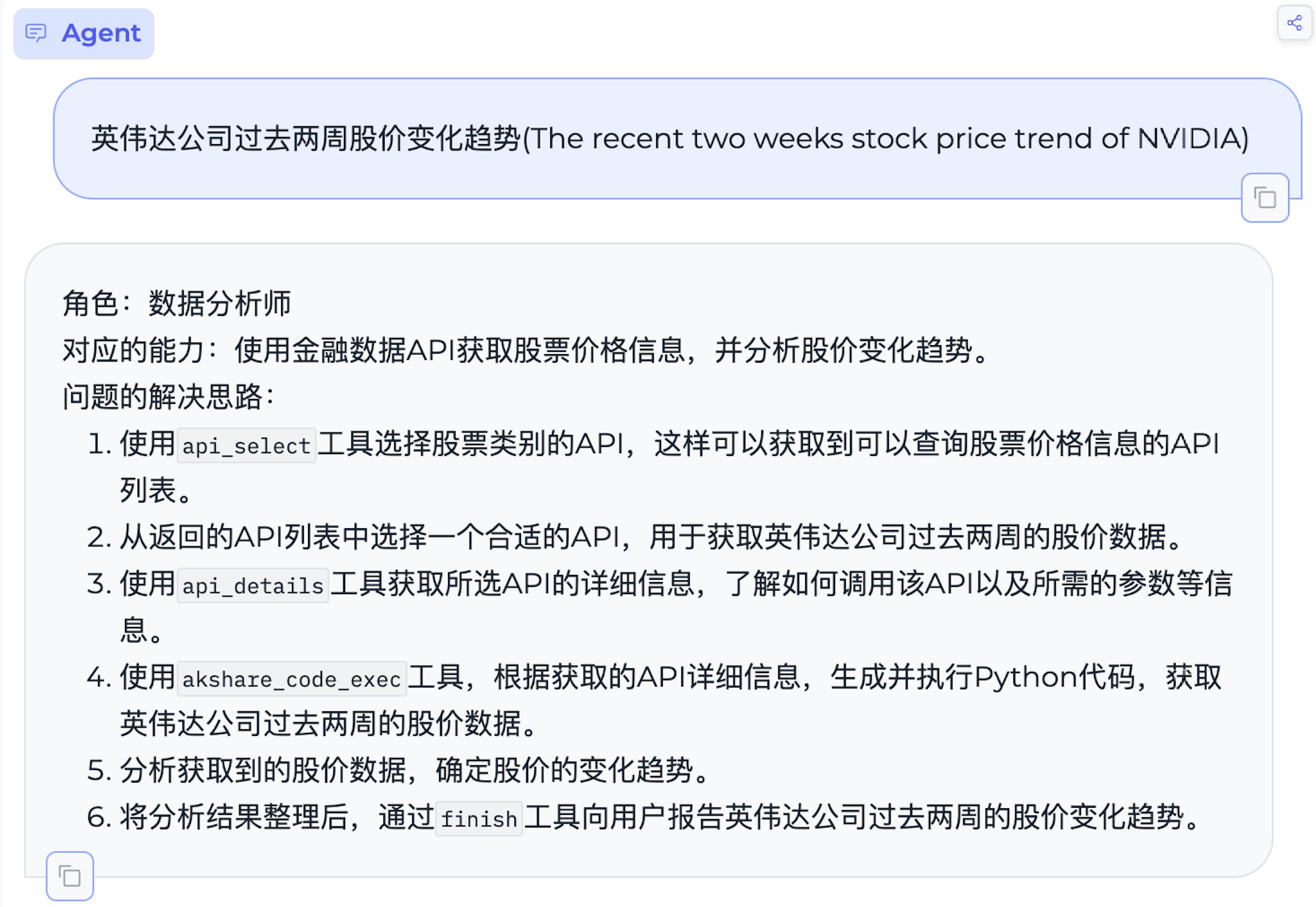}
    \centering
    \caption{Demo showcase of agent meta}
    \label{fig:minipage1}
\end{figure}
\begin{figure}[h]
    \centering
    \includegraphics[width=1\linewidth]{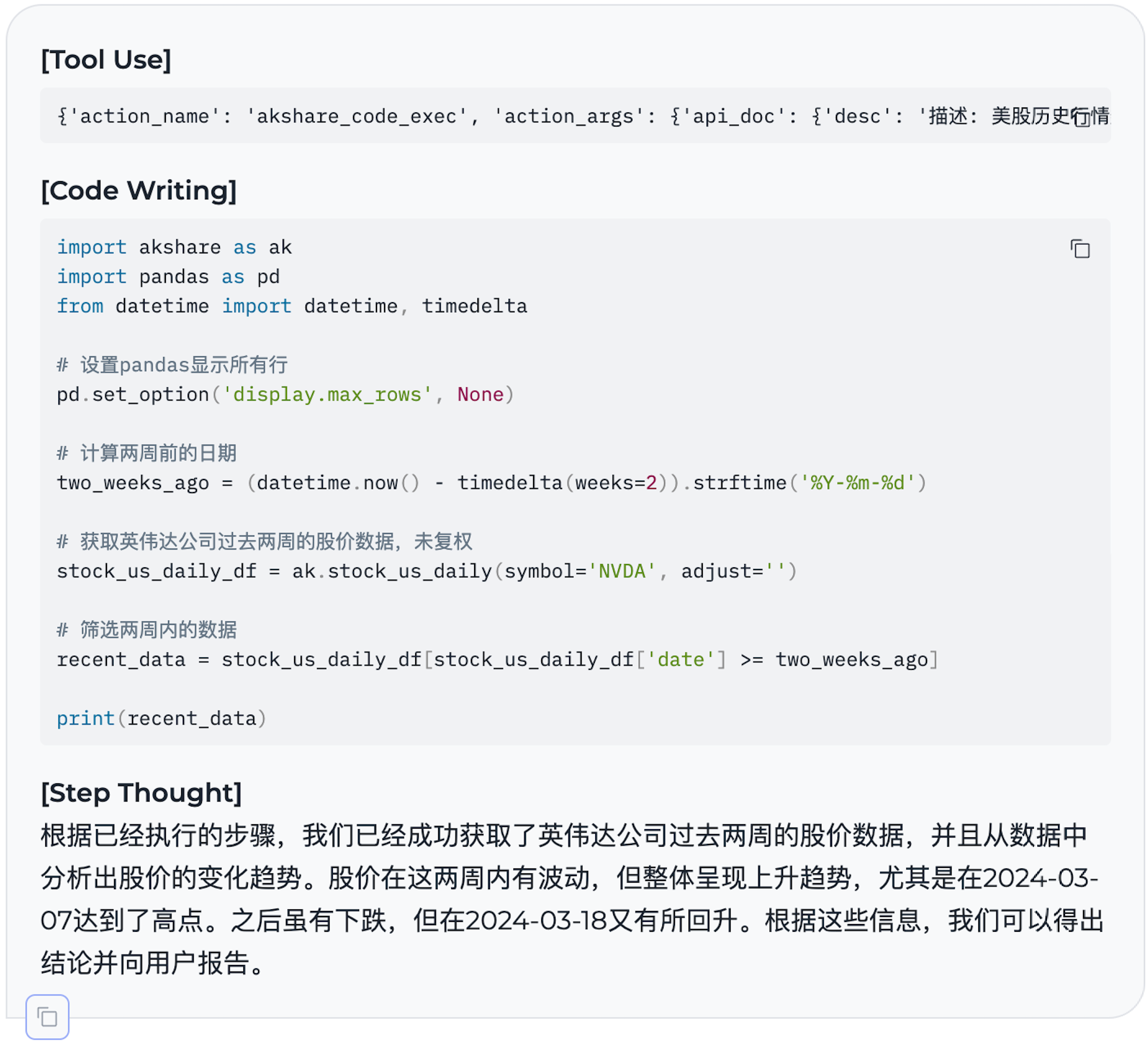}
    \centering
    \caption{Demo showcase of code-exec}
    \label{fig:minipage2}
\end{figure}

In Fig.\ref{fig:minipage1} and Fig.\ref{fig:minipage2}, we present two steps of a demo that analyzes the stock price trend of NVIDIA Corporation. The prompts of the two steps could be found in appendix. When the agent receives a query from the user, it determines roles and capabilities based on this query, which guides the subsequent steps in accordance with the Overall Plan. In this case, the LLMs employs a combination of three actions: \textit{api-select}, \textit{api-details} and \textit{code-exec} to compose the code for data retrieval and analysis. Post-execution of the code, the LLMs undertake the task of analyzing and summarizing the results. Once the model assesses that the task has been successfully resolved, it concludes the process with \textit{finish} action and provides a conclusion. Due to the constraints of the paper's length, for a more concrete demonstration, please refer to the video we uploaded on YouTube.

\section{Conclusion}
In this paper, we introduce FinVerse, an agent system tailored to financial domains. FinVerse is furnished with over 600 specialized APIs for accessing real-time financial data and possesses an integrated code interpreter for executing sophisticated data analysis tasks. Moreover, we have augmented LLMs for FinVerse with SFT to enhance primarily four capabilities. Experimental evaluations on a well-defined dataset show that FinVerse can achieve impressive performance with satisfactory resource efficiency.




\section*{Ethics Statement}
The authors declare that the data in our work is publicly available and does not involve political and moral sensitivities. 
Ethical concerns include the usage of the proposed solution for a purpose different from that previously mentioned in the paper, such as the inputs of racism, violence, etc.

\bibliography{anthology,custom}

\begin{thebibliography}{26}
\expandafter\ifx\csname natexlab\endcsname\relax\def\natexlab#1{#1}\fi

\bibitem[{Bai et~al.(2023)Bai, Bai, Chu, Cui, Dang, Deng, Fan, Ge, Han, Huang et~al.}]{bai2023qwen}
Jinze Bai, Shuai Bai, Yunfei Chu, Zeyu Cui, Kai Dang, Xiaodong Deng, Yang Fan, Wenbin Ge, Yu~Han, Fei Huang, et~al. 2023.
\newblock Qwen technical report.
\newblock \emph{arXiv preprint arXiv:2309.16609}.

\bibitem[{Baumann et~al.(2024)Baumann, Diaz, Michael, Netz, Nqiri, Reimer, and Rumpe}]{baumann2024combining}
Nils Baumann, Juan~Sebastian Diaz, Judith Michael, Lukas Netz, Haron Nqiri, Jan Reimer, and Bernhard Rumpe. 2024.
\newblock Combining retrieval-augmented generation and few-shot learning for model synthesis of uncommon dsls.
\newblock In \emph{Modellierung 2024 Satellite Events}, pages 10--18420. Gesellschaft f{\"u}r Informatik eV.

\bibitem[{Brown et~al.(2020)Brown, Mann, Ryder, Subbiah, Kaplan, Dhariwal, Neelakantan, Shyam, Sastry, Askell et~al.}]{brown2020language}
Tom Brown, Benjamin Mann, Nick Ryder, Melanie Subbiah, Jared~D Kaplan, Prafulla Dhariwal, Arvind Neelakantan, Pranav Shyam, Girish Sastry, Amanda Askell, et~al. 2020.
\newblock Language models are few-shot learners.
\newblock \emph{Advances in neural information processing systems}, 33:1877--1901.

\bibitem[{Chen et~al.(2024)Chen, Xiao, Zhang, Luo, Lian, and Liu}]{chen2024bge}
Jianlv Chen, Shitao Xiao, Peitian Zhang, Kun Luo, Defu Lian, and Zheng Liu. 2024.
\newblock \href {http://arxiv.org/abs/2402.03216} {Bge m3-embedding: Multi-lingual, multi-functionality, multi-granularity text embeddings through self-knowledge distillation}.

\bibitem[{Chen et~al.(2023{\natexlab{a}})Chen, Lin, Han, and Sun}]{chen2023benchmarking}
Jiawei Chen, Hongyu Lin, Xianpei Han, and Le~Sun. 2023{\natexlab{a}}.
\newblock \href {http://arxiv.org/abs/2309.01431} {Benchmarking large language models in retrieval-augmented generation}.

\bibitem[{Chen et~al.(2023{\natexlab{b}})Chen, Su, Zuo, Yang, Yuan, Chan, Yu, Lu, Hung, Qian, Qin, Cong, Xie, Liu, Sun, and Zhou}]{chen2023agentverse}
Weize Chen, Yusheng Su, Jingwei Zuo, Cheng Yang, Chenfei Yuan, Chi-Min Chan, Heyang Yu, Yaxi Lu, Yi-Hsin Hung, Chen Qian, Yujia Qin, Xin Cong, Ruobing Xie, Zhiyuan Liu, Maosong Sun, and Jie Zhou. 2023{\natexlab{b}}.
\newblock \href {http://arxiv.org/abs/2308.10848} {Agentverse: Facilitating multi-agent collaboration and exploring emergent behaviors}.

\bibitem[{Gao et~al.(2024)Gao, Xiong, Gao, Jia, Pan, Bi, Dai, Sun, Guo, Wang, and Wang}]{gao2024retrievalaugmented}
Yunfan Gao, Yun Xiong, Xinyu Gao, Kangxiang Jia, Jinliu Pan, Yuxi Bi, Yi~Dai, Jiawei Sun, Qianyu Guo, Meng Wang, and Haofen Wang. 2024.
\newblock \href {http://arxiv.org/abs/2312.10997} {Retrieval-augmented generation for large language models: A survey}.

\bibitem[{Hatalis et~al.(2023)Hatalis, Christou, Myers, Jones, Lambert, Amos-Binks, Dannenhauer, and Dannenhauer}]{hatalis2023memory}
Kostas Hatalis, Despina Christou, Joshua Myers, Steven Jones, Keith Lambert, Adam Amos-Binks, Zohreh Dannenhauer, and Dustin Dannenhauer. 2023.
\newblock Memory matters: The need to improve long-term memory in llm-agents.
\newblock In \emph{Proceedings of the AAAI Symposium Series}, volume~2, pages 277--280.

\bibitem[{Hong et~al.(2023)Hong, Zheng, Chen, Cheng, Wang, Zhang, Wang, Yau, Lin, Zhou et~al.}]{hong2023metagpt}
Sirui Hong, Xiawu Zheng, Jonathan Chen, Yuheng Cheng, Jinlin Wang, Ceyao Zhang, Zili Wang, Steven Ka~Shing Yau, Zijuan Lin, Liyang Zhou, et~al. 2023.
\newblock Metagpt: Meta programming for multi-agent collaborative framework.
\newblock \emph{arXiv preprint arXiv:2308.00352}.

\bibitem[{Nakano et~al.(2022)Nakano, Hilton, Balaji, Wu, Ouyang, Kim, Hesse, Jain, Kosaraju, Saunders, Jiang, Cobbe, Eloundou, Krueger, Button, Knight, Chess, and Schulman}]{nakano2022webgpt}
Reiichiro Nakano, Jacob Hilton, Suchir Balaji, Jeff Wu, Long Ouyang, Christina Kim, Christopher Hesse, Shantanu Jain, Vineet Kosaraju, William Saunders, Xu~Jiang, Karl Cobbe, Tyna Eloundou, Gretchen Krueger, Kevin Button, Matthew Knight, Benjamin Chess, and John Schulman. 2022.
\newblock \href {http://arxiv.org/abs/2112.09332} {Webgpt: Browser-assisted question-answering with human feedback}.

\bibitem[{Ouyang et~al.(2022)Ouyang, Wu, Jiang, Almeida, Wainwright, Mishkin, Zhang, Agarwal, Slama, Ray, Schulman, Hilton, Kelton, Miller, Simens, Askell, Welinder, Christiano, Leike, and Lowe}]{ouyang2022training}
Long Ouyang, Jeff Wu, Xu~Jiang, Diogo Almeida, Carroll~L. Wainwright, Pamela Mishkin, Chong Zhang, Sandhini Agarwal, Katarina Slama, Alex Ray, John Schulman, Jacob Hilton, Fraser Kelton, Luke Miller, Maddie Simens, Amanda Askell, Peter Welinder, Paul Christiano, Jan Leike, and Ryan Lowe. 2022.
\newblock \href {http://arxiv.org/abs/2203.02155} {Training language models to follow instructions with human feedback}.

\bibitem[{Pan et~al.(2024)Pan, Zhai, Yuan, Lv, Fu, Liu, Wang, and Qin}]{pan2024kwaiagents}
Haojie Pan, Zepeng Zhai, Hao Yuan, Yaojia Lv, Ruiji Fu, Ming Liu, Zhongyuan Wang, and Bing Qin. 2024.
\newblock \href {http://arxiv.org/abs/2312.04889} {Kwaiagents: Generalized information-seeking agent system with large language models}.

\bibitem[{Qin et~al.(2023{\natexlab{a}})Qin, Cai, Jin, Yan, Liang, Zhu, Lin, Han, Ding, Wang, Xie, Qi, Liu, Sun, and Zhou}]{qin2023webcpm}
Yujia Qin, Zihan Cai, Dian Jin, Lan Yan, Shihao Liang, Kunlun Zhu, Yankai Lin, Xu~Han, Ning Ding, Huadong Wang, Ruobing Xie, Fanchao Qi, Zhiyuan Liu, Maosong Sun, and Jie Zhou. 2023{\natexlab{a}}.
\newblock \href {http://arxiv.org/abs/2305.06849} {Webcpm: Interactive web search for chinese long-form question answering}.

\bibitem[{Qin et~al.(2023{\natexlab{b}})Qin, Liang, Ye, Zhu, Yan, Lu, Lin, Cong, Tang, Qian, Zhao, Hong, Tian, Xie, Zhou, Gerstein, Li, Liu, and Sun}]{qin2023toolllm}
Yujia Qin, Shihao Liang, Yining Ye, Kunlun Zhu, Lan Yan, Yaxi Lu, Yankai Lin, Xin Cong, Xiangru Tang, Bill Qian, Sihan Zhao, Lauren Hong, Runchu Tian, Ruobing Xie, Jie Zhou, Mark Gerstein, Dahai Li, Zhiyuan Liu, and Maosong Sun. 2023{\natexlab{b}}.
\newblock \href {http://arxiv.org/abs/2307.16789} {Toolllm: Facilitating large language models to master 16000+ real-world apis}.

\bibitem[{Schick et~al.(2023)Schick, Dwivedi-Yu, Dessì, Raileanu, Lomeli, Zettlemoyer, Cancedda, and Scialom}]{schick2023toolformer}
Timo Schick, Jane Dwivedi-Yu, Roberto Dessì, Roberta Raileanu, Maria Lomeli, Luke Zettlemoyer, Nicola Cancedda, and Thomas Scialom. 2023.
\newblock \href {http://arxiv.org/abs/2302.04761} {Toolformer: Language models can teach themselves to use tools}.

\bibitem[{Shinn et~al.(2023)Shinn, Cassano, Berman, Gopinath, Narasimhan, and Yao}]{shinn2023reflexion}
Noah Shinn, Federico Cassano, Edward Berman, Ashwin Gopinath, Karthik Narasimhan, and Shunyu Yao. 2023.
\newblock \href {http://arxiv.org/abs/2303.11366} {Reflexion: Language agents with verbal reinforcement learning}.

\bibitem[{Talebirad and Nadiri(2023)}]{talebirad2023multi}
Yashar Talebirad and Amirhossein Nadiri. 2023.
\newblock Multi-agent collaboration: Harnessing the power of intelligent llm agents.
\newblock \emph{arXiv preprint arXiv:2306.03314}.

\bibitem[{Team(2023)}]{xagent2023}
XAgent Team. 2023.
\newblock Xagent: An autonomous agent for complex task solving.

\bibitem[{Touvron et~al.(2023)Touvron, Martin, Stone, Albert, Almahairi, Babaei, Bashlykov, Batra, Bhargava, Bhosale et~al.}]{touvron2023llama}
Hugo Touvron, Louis Martin, Kevin Stone, Peter Albert, Amjad Almahairi, Yasmine Babaei, Nikolay Bashlykov, Soumya Batra, Prajjwal Bhargava, Shruti Bhosale, et~al. 2023.
\newblock Llama 2: Open foundation and fine-tuned chat models.
\newblock \emph{arXiv preprint arXiv:2307.09288}.

\bibitem[{Wang et~al.(2023)Wang, Ma, Feng, Zhang, Yang, Zhang, Chen, Tang, Chen, Lin et~al.}]{wang2023survey}
Lei Wang, Chen Ma, Xueyang Feng, Zeyu Zhang, Hao Yang, Jingsen Zhang, Zhiyuan Chen, Jiakai Tang, Xu~Chen, Yankai Lin, et~al. 2023.
\newblock A survey on large language model based autonomous agents.
\newblock \emph{arXiv preprint arXiv:2308.11432}.

\bibitem[{Wu et~al.(2023)Wu, Bansal, Zhang, Wu, Zhang, Zhu, Li, Jiang, Zhang, and Wang}]{wu2023autogen}
Qingyun Wu, Gagan Bansal, Jieyu Zhang, Yiran Wu, Shaokun Zhang, Erkang Zhu, Beibin Li, Li~Jiang, Xiaoyun Zhang, and Chi Wang. 2023.
\newblock Autogen: Enabling next-gen llm applications via multi-agent conversation framework.
\newblock \emph{arXiv preprint arXiv:2308.08155}.

\bibitem[{Xie et~al.(2024)Xie, Zhang, Chen, Zhu, Lou, Tian, Xiao, and Su}]{xie2024travelplanner}
Jian Xie, Kai Zhang, Jiangjie Chen, Tinghui Zhu, Renze Lou, Yuandong Tian, Yanghua Xiao, and Yu~Su. 2024.
\newblock Travelplanner: A benchmark for real-world planning with language agents.
\newblock \emph{arXiv preprint arXiv:2402.01622}.

\bibitem[{Yang et~al.(2023)Yang, Xiao, Wang, Zhang, Bian, Yin, Lv, Pan, Wang, Yan et~al.}]{yang2023baichuan}
Aiyuan Yang, Bin Xiao, Bingning Wang, Borong Zhang, Ce~Bian, Chao Yin, Chenxu Lv, Da~Pan, Dian Wang, Dong Yan, et~al. 2023.
\newblock Baichuan 2: Open large-scale language models.
\newblock \emph{arXiv preprint arXiv:2309.10305}.

\bibitem[{Yao et~al.(2022)Yao, Zhao, Yu, Du, Shafran, Narasimhan, and Cao}]{yao2022react}
Shunyu Yao, Jeffrey Zhao, Dian Yu, Nan Du, Izhak Shafran, Karthik Narasimhan, and Yuan Cao. 2022.
\newblock React: Synergizing reasoning and acting in language models.
\newblock \emph{arXiv preprint arXiv:2210.03629}.

\bibitem[{Zeng et~al.(2023)Zeng, Liu, Lu, Wang, Liu, Dong, and Tang}]{zeng2023agenttuning}
Aohan Zeng, Mingdao Liu, Rui Lu, Bowen Wang, Xiao Liu, Yuxiao Dong, and Jie Tang. 2023.
\newblock \href {http://arxiv.org/abs/2310.12823} {Agenttuning: Enabling generalized agent abilities for llms}.

\bibitem[{Zhang et~al.(2023)Zhang, Shen, Lu, and Zhuang}]{zhang2023datacopilot}
Wenqi Zhang, Yongliang Shen, Weiming Lu, and Yueting Zhuang. 2023.
\newblock \href {http://arxiv.org/abs/2306.07209} {Data-copilot: Bridging billions of data and humans with autonomous workflow}.

\end{thebibliography}
\bibliographystyle{acl_natbib}

\appendix


\section{Related Works}

The open-source community has put forth a variety of versatile agent frameworks, notable generalist examples being AutoGPT and XAgent \citep{xagent2023}. Platforms such as AgentVerse \citep{chen2023agentverse} and MetaGPT \citep{hong2023metagpt} offer collaborative solutions for the operation of multiple agents. Data-Copilot \citep{zhang2023datacopilot} introduces a method for creating data analysis-centric agent applications, incorporating intent recognition and pipeline construction, subsequently enabling scenarios like stock query services. Furthermore, KwaiAgent \citep{pan2024kwaiagents} delivers a comprehensive framework for generic agents, complete with evaluation benchmarks. It has been specifically fine-tuned on open-source models to enhance its integration into its agent architecture.

\section{Details of plugins}
\label{sec:dataset details}
Table\,\ref{tab:tools description} presents the specific categories of tools utilized by FinVerse along with the corresponding types of plugins for each category. The general tools include a website search, a code interpreter, and finish symbol. Here, we provide a sandboxed environment for the execution of Python code, where we offer third-party libraries such as Matplotlib, Reportlab, Pandas, and Numpy for FinVerse's use. Moreover, within the sandbox, we have installed Akshares as specialized tools and compiled a selection of API documentation to serve as our candidate API set.
Finally, FinVerse has access to financial APIs spanning ten categories, as listed in Table\,\ref{tab:tools description}, with the Stock and Macroeconomics categories being the most extensive. When making selections, vector search is employed to manage the size of the candidate set for Stock and Macroeconomics.

\begin{table}[h]
\scalebox{1.0}
{
\begin{tabular}{lll}
\toprule
Categories & Name & Diversity   \\
       \midrule
General & Web Search  & 3                  \\
        & Code Interpreter & 1 \\
        & Finish & 1 \\
       \midrule
Specific & Stock  & 243                                           \\
&Fund   & 46                                           \\
&Futures& 35  \\
&Foreign exchange  & 6  \\
&Index  & 59  \\
&Interest Rate  & 9   \\
&Currency   & 3   \\
&Macroeconomics & 186 \\
&Option & 32 \\
&Bond & 23 \\
\bottomrule
\end{tabular}
}
\caption{The basic statistics of tools in FinVerse}
\label{tab:tools description}
\end{table}

\section{Training Hyperparameters}
In Table \ref{tab:hyperparameters of llms}, we provide the detailed configurations used in Sparse Fine-Tuning (SFT) when training the Baichuan2-13b-chat model and the Qwen-14b-chat model, to facilitate the reproduction of our experimental results. For both foundational models, we employ an output length of 8k during training to ensure sufficient memory space for multi-step decision-making required within FinVerse. Additionally, both models are trained and perform inference on NVIDIA's A100 graphics cards.

\begin{table}[h]
\small\centering
  \label{Hyperparameters of UniVKIE}
  \resizebox{0.45\textwidth}{!}{
  \begin{tabular}{lll}
    \toprule
Hyperparameters & FV-BC2 & FV-QW\\
    \midrule
    max source length & 8192 & 8192 \\
    max target length & 2048  & 2048\\
    gradient checkpoint & 1 & 1 \\
    batch size & 32 & 16 \\  
    learning rate & 5e-5 & 1e-5\\
    training epochs & 2 & 2\\
    precision format & bf16 & bf16\\
    lr scheduler & cosine & cosine\\
    warmup steps & 1500 & 1000\\
    weight decay & 0.1 & 0.1\\
    Adam beta1 & 0.9 & 0.9\\
    Adam beta2 & 0.95 & 0.95\\
    Deepspeed & zero3 & zero3\\
    \bottomrule
\end{tabular}}
\caption{SFT Hyperparameters of LLMs}
\label{tab:hyperparameters of llms}
\end{table}

\section{Prompt Templetes}
\label{sec:platform}
\begin{figure*}[h]

    \centering
    \includegraphics[width=1\linewidth]{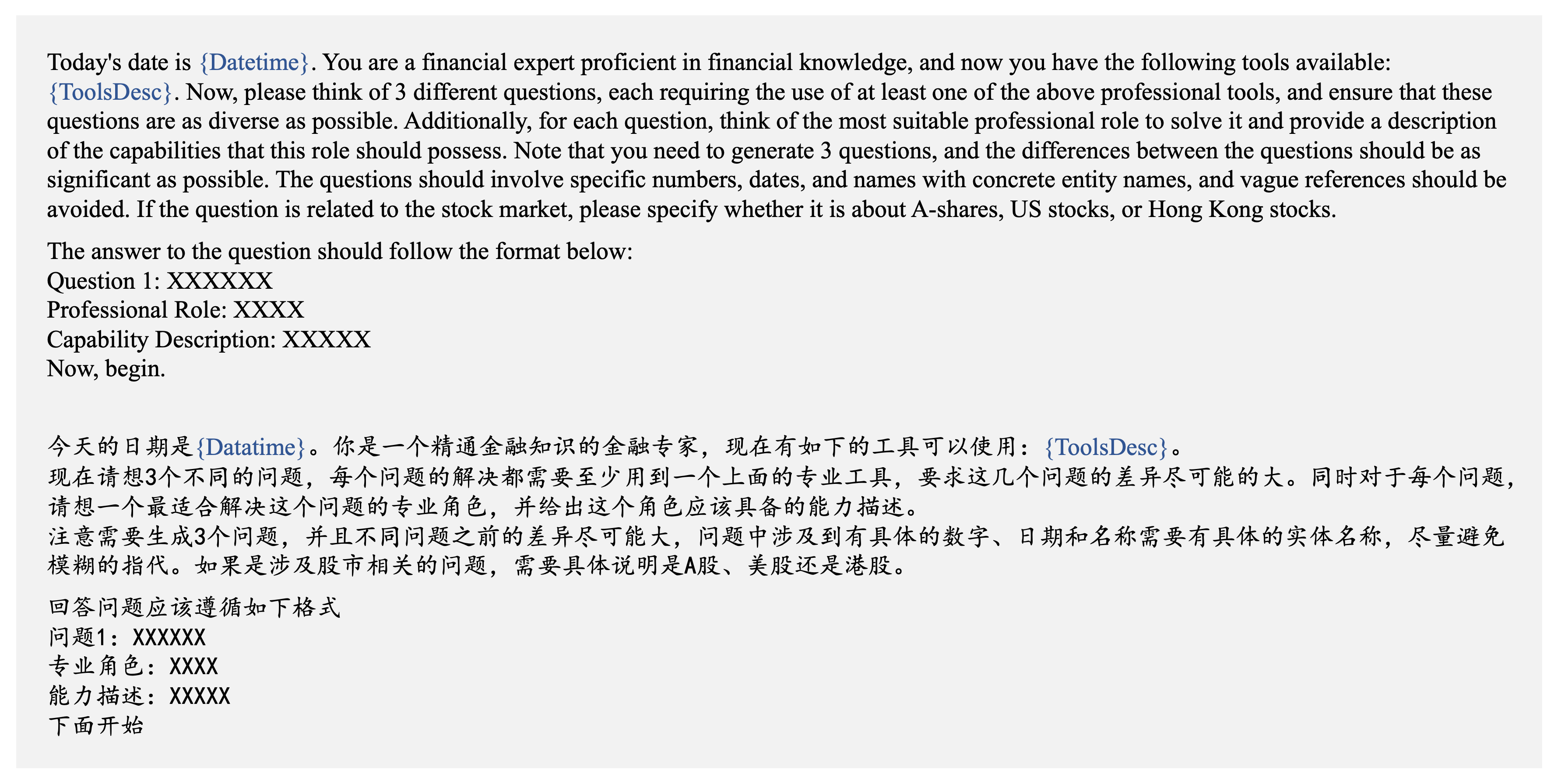}
    \centering
    \caption{Prompt to generate agent meta.}
    \label{fig:plan-roles}
\end{figure*}

\begin{figure*}[h]
    \centering
    \includegraphics[width=1\linewidth]{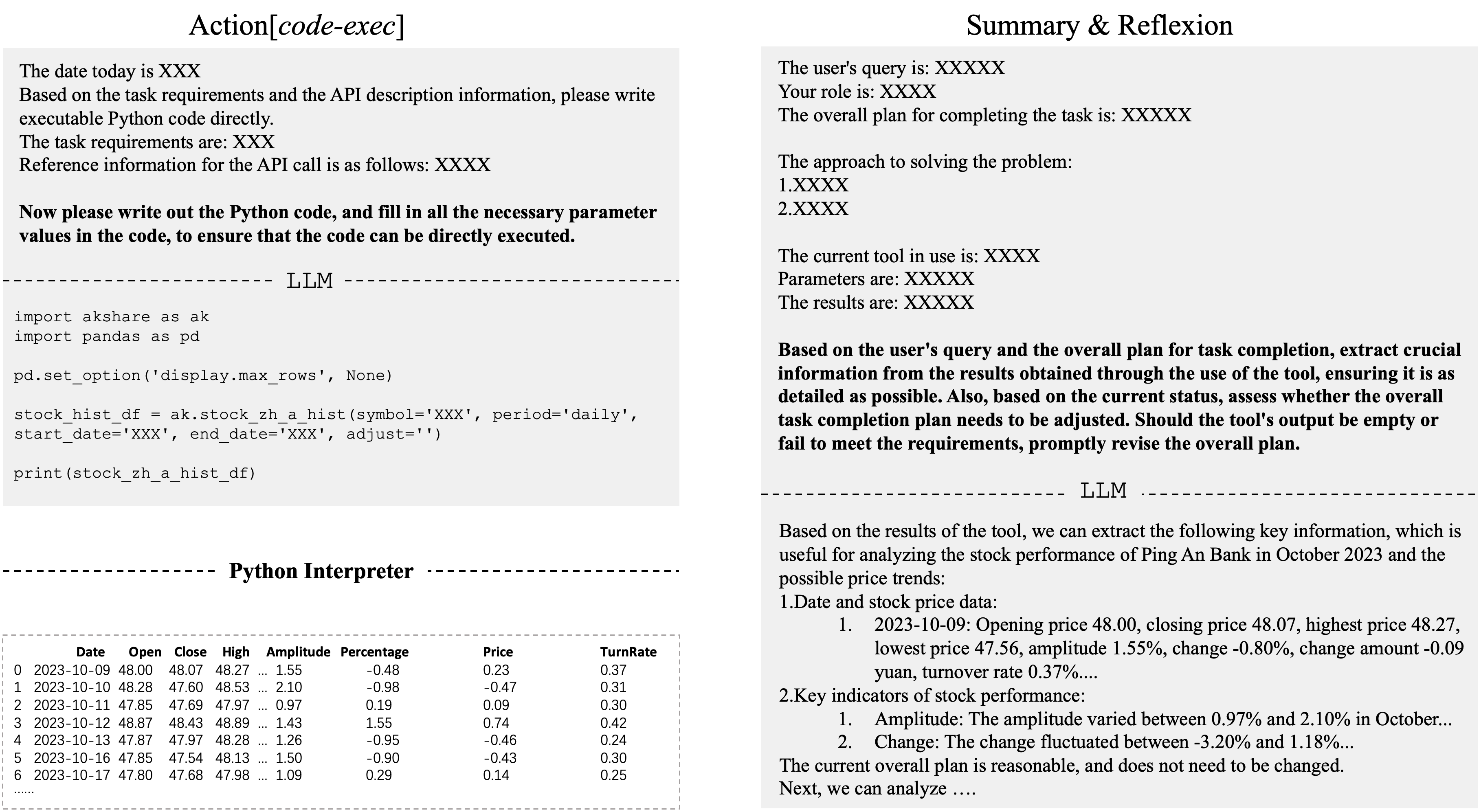}
    \centering
    \caption{Prompts corresponding to code-exec and summary\&reflexion }.
    \label{fig:code-exec}
\end{figure*}

In this section, we offer a selection of prompt templates used in our system construction for reference as in Fig.\ref{fig:plan-roles} and Fig.\ref{fig:code-exec}.

.

\end{document}